\documentclass[epj]{svjour}
\usepackage{graphicx,amsfonts}

\begin{document}

\title{Fragmentation of a circular disc by projectiles}
\author{Bhupalendra Behera\inst{1,3} \and
 Ferenc Kun\inst{2} \and Sean McNamara\inst{1} \and Hans J.\ Herrmann\inst{1}}
\institute{Institut f\"ur Computeranwendungen (ICA1), Universit\"at
  Stuttgart, 70569 Stuttgart, Germany \and 
  Department of Theoretical Physics, University of
  Debrecen, H-4010 Debrecen, P.O.Box: 5, Hungary \and
  Indian Institute of Technology, Department of Materials 
  and Metallurgical Engineering, Kanpur, India
}

\date{Received: date / Revised version: date}

\offprints{bhupalendra@yahoo.com}

\abstract{
The fragmentation of a two-dimensional circular disc  by lateral
impact is
investigated using a cell model of brittle solid. The disc is composed of
numerous unbreakable randomly shaped convex polygons connected  together by
simple elastic beams that break when bent or stretched beyond a certain
limit. 
We found that the fragment mass distribution follows
a power law with an exponent close to $2$ independent of the system size.
We also observed two types of crack patterns: radial cracks starting
from the impact point and cracks perpendicular to the radial ones. 
Simulations revealed that there exists a critical projectile energy, above
which the target breaks into numerous smaller pieces, and below
which it suffers only damage in the form of cracks. Our theoretical
results are in a reasonable agreement with recent experimental
findings on the fragmentation of discs.
}
\maketitle

\section{Introduction}
\label{intro}

Fragmentation of finite size systems is a widespread phenomenon in nature
which is of considerable scientific and industrial interest. Despite the
intense research in various fields of science and technology, complete
analysis and understanding of fragmentation has not yet been
achieved. 
In industry
comminution, or the process of 
size reduction of granular materials, is very important.
There is particular interest in the net energy required to achieve a
certain size reduction and the energy distribution of fragments during the
fragmentation process. For different comminution operations like crushing
and grinding, early laws were developed by Von Rittinger, Kick and Bond
\cite{mineral}, having a relatively narrow size range of
applicability. 
Many experimental and simulational studies have been undertaken to
extend our understanding of fragmentation to a wider range of situations.

Fragmentation occurs over a wide range, from the collisional evolution of
asteroids \cite{asteroids} to the fragmentation of mo\-le\-cu\-les
\cite{critical,cluster}. 
At intermediate scales, there is
the degradation of materials comprising small agglomerates employed in
process
industries such as pharmaceuticals, chemicals, fertilizers and detergents.
There are also many geological examples associated
with the use of explosives for mining such as oil shale industry, coal
heaps etc.  It was found in many cases that the probability density
function $F(m)$ of fragment masses follows a power law
\begin{equation}
F(m) \sim m^{-\tau}.
\end{equation}

Several numerical 
\cite{cluster,diehl,agglomerate,discrete,transition,twodisc,granulate,britt,droplet,aspect,branching,univer}
and experimental \cite{asteroids,ice,eggs,composite,proton,self,scaling,platelike,glassplate} studies
have been performed on the fragmentation of solids and particle
agglomerates induced by impact
under different conditions. In all of these studies one obtains
$F(m) \sim m^{-\tau}$ for the fragment mass distribution with an
exponent $\tau$ depending on the 
dimensionality of the system, {\it i.e.} in higher dimensions
usually a larger value of $\tau$ is obtained 
\cite{asteroids,critical,cluster,diehl,agglomerate,discrete,transition,twodisc,granulate,britt,droplet,aspect,branching,univer,ice,eggs,composite,proton,self,scaling,platelike,glassplate}. 
Table \ref{numericaltab} summarizes the most important numerical and
experimental results on impact fragmentation. 

Recently, ``lateral experiments'' where a hypervelocity projectile strikes
the side of a plate was carried out by T.\ Kadono \cite{platelike}.
Moreover, another experiment was carried by T.\ Kadono and
M.\ Arakawa \cite{glassplate} by shooting a cylindrical aluminum
projectile with a diameter of 15mm and height of 10mm on thin Pyrex
glass plate targets.  They calculated the cumulative number of
fragments larger than 
a given mass, and found that the resulting function was a power law
with an exponent of about $0.60$.  This function is the integral of
$F(m)$, so their results correspond to $\tau = 1.60$. They found two 
types of crack pattern during fragmentation process: radial cracks 
initiating from the impact point and cracks perpendicular to the 
radial ones. They considered that the radial cracks formed prior to the 
perpendicular ones \cite{platelike,glassplate}.
\begin{table}
\caption{\label{numericaltab}
A Summary of recent numerical and experimental studies of fragmentation,
showing the exponent of the fragment mass distribution.}
\begin{tabular}{lc}
\hline\noalign{\smallskip}
{\bf Numerical Results} & {\bf Exponent $\tau$ }\\
\hline
Explosion of disc-shaped solids \cite{discrete,granulate} & 2.0\\
Impact of a projectile into a block \cite{discrete,granulate} &
1.98\\
Collision of two discs \cite{transition,twodisc} & 1.75 - 2.27\\
Lennard-Jones solid in d=2  \cite{diehl,univer,ching}  & 1.4-1.5 \\
\hline
{\bf Experimental Results} & {\bf Exponent $\tau$}\\
\hline
Collisional disruption of ice \cite{ice} & $1.64 \pm 0.06$ \\
Fragmentation of shells \cite{eggs}       & 1.35   \\
Fragmentation of dry clay plates         & \\
(for large fragments) \cite{composite,self}   & 1.12 - 1.27 \\
Fragmentation of dry clay plates         & \\
(for small fragments) \cite{composite}   & 1.5 - 1.67 \\
Proton-nucleus collision \cite{proton}   & 2.6 \\
Free fall of glass rods  \cite{glassrod} & 1.2-1.8 \\
Impact on glass tubes \cite{scaling}     & 1.5 \\
Fragmentation of plaster plates \cite{platelike} & 1.1 - 1.3 \\
Fragmentation of glass plates \cite{glassplate} & 1.7 \\
Peripheral Au + Au collisions \cite{Au}  & 2.2 \\
\noalign{\smallskip}\hline
\end{tabular}
\end{table}

In the present paper we have studied the lateral impact of a
projectile into the right hand side of a disk following the
experiments of Refs.\ \cite{platelike,glassplate}. Here we have taken
a particle composed of numerous unbreakable, undeformable, randomly shaped
polygons which are bonded together by elastic beams. This model has been
used before for the study of the fragmentation process in different
physical
situations \cite{discrete,transition,twodisc,granulate}. The contacts
between the elements (polygons) can be broken according to a
physical breaking 
rule, which takes into account the stretching and bending of the
connections.
Based on simulations of the model, we performed a detailed study of the
failure evolution at different impact energies 
and of the nature of the crack propagation during the fragmentation
process.
We also study the statistics of average fragment mass and average largest
mass, the energy and velocity distributions of fragments during the
fragmentation process when changing the imparted energy.

We have given an outline of the theoretical background of the model in
Sec.~\ref{model}, a short description of the basic mechanism of
fragmentation process and crack propagation is presented in
Sec.~\ref{crack}. The numerical results at different 
initial conditions
are given in Sec.~\ref{results}, where the velocity, energy and
mass distribution of fragments, and the transition from damage to
fragmentation are studied.

\section{The model}
\label{model}

To study fragmentation of granular solids, we performed molecular dynamic
simulations (MD) in two dimensions. In order to better capture the complex
structure of real solids, we used arbitrarily shaped convex polygons
that interact with each other elastically. The model consists of three
major parts, namely, the construction of a Voronoi cellular structure, the
introduction of the elastic behavior, and finally the breaking of the solid.

The Voronoi construction, which is a random tessellation of the plane
into convex polygons, was used to reflect the grain structure of the solid.
One puts a random set
of points onto the plane and then assigns each point that part of the
plane which is nearer to it than to any other point. In our case,
the initial configuration of the polygons was constructed using a vectorizable
random lattice, which is a Voronoi construction with reduced
disorder \cite{rand_lattic}. One advantage of the Voronoi tessellation is
that the number of neighbors of each polygon is limited which makes the
computer code faster and allows us to simulate larger systems.
We generate a square block of
Voronoi cellular structure from which we cut out a disk,
special care being taken to get a smooth outer surface.
With different seed values of the Voronoi generator, different
samples are obtained with differently shaped Voronoi cells.
Previously this model has been applied to study fragmentation of solids in
various experimental situations
\cite{discrete,transition,twodisc,granulate}. 

In the model the polygons are rigid bodies with 
three degrees of freedom in two dimensions: the two
coordinates of the center of mass and the rotation angle.
They are neither breakable nor
deformable but they can overlap when pressed against each other. The
overlap represents local deformation of the grains. Usually the
overlapping polygons have two intersection points which define the
contact line. In order to simulate the elastic contact force, we introduce
a repulsive force between touching polygons. This force is proportional to
the overlapping area $A$ divided by a characteristic length $L_c$
($\frac{1}{L_c}= \frac{1}{2}(\frac{1}{r_i}+\frac{1}{r_j})$, where
$r_i$, $r_j$ are the radii of circles of the same area of polygons),
multiplied by a spring constant that 
is proportional to the elastic modulus divided by the
characteristic length. The direction of the force is chosen to be
perpendicular to the contact line of the polygons. Further, damping and
friction of the touching polygons according to Coulomb's friction law are
also implemented.

To bond the particles together, the centers of mass of neighboring
polygons are joined together with beams that exert an attractive,
restoring force but can break in order to model the
fragmentation of the solid. The cross section of the beams is the length
of the common side of the neighboring polygons and the length is the
distance between the centers of mass of the two polygons. The Young's
modulus of the beams $E_b$ and of the particles $E_p$ are independent of
each other. The beams can be broken according to a physical breaking rule
of the form of the von Mises plasticity criterion, which takes into
account the stretching and bending of the connection
\begin{equation}
P_b^{ij}=\left(\frac{\epsilon_b^{ij}}{\epsilon_ {b,\mathrm{max}}}\right)^2
+ \frac{\mathrm{max}(|\theta^i|,|\theta^j|)}{\theta_\mathrm{max}} \ge 1,
\quad \epsilon_b^{ij} \ge 0,
\label{BreakingCondition}
\end{equation}
where $\epsilon_b^{ij}=\Delta\ell^{ij}/\ell^{ij}$ is the longitudinal
strain of the beam, $\theta^i$ and $\theta^j$ are the rotation angles at
the two ends of the beams, and $\epsilon_{b,max}$ and $\theta_\mathrm{max}$
are the threshold value of the two breaking modes. The breaking condition
is checked at each iteration time step and those beams where condition
(\ref{BreakingCondition})
holds are removed from the calculation, and never restored. The surface of
the grains on which beams are broken represents cracks. The energy stored
in the broken beams represents the energy needed to create these new crack
surfaces inside the solid. 

The time evolution of the system is obtained by numerically solving
Newton's equations of motion of the individual polygons (Molecular
Dynamics).  For the solution of the equations we use a Gear
Predictor-Corrector scheme of fifth order, which means that we have to
keep track of the coordinates and all their derivatives up to fifth order.

\begin{table}
\begin{tabular}{lccc}
\hline
{\bf Parameters} & {\bf symbol} & {\bf Unit} & {\bf Value}\\
\hline
Density & $ \rho$ & $ g/cm^3$ & $5$ \\
Grain Young's modulus & $E_p$ & $ dyne/cm^2$ & $10^{10}$\\
Beams Young's modulus & $E_b$ & $dyne/cm^2$ & $10^9$\\
Time step & $dt$ & $s$ & $10^{-7}$ \\
Failure elongation & $\epsilon_{b,max}$ & $\%$ & 3 \\
Failure bending  & $\theta_{b,\mathrm{max}}$ & degree & $ 1$ \\
\hline
\end{tabular}
\caption{\label{simulvalue}
 The parameter values used in the simulations.}
\end{table}

\section{Fragmentation process and Crack Propagation}
\label{crack}

In the present paper we apply our model to explore the properties of
the fragmentation process of a circular disc due to an
impacting projectile.   
On the right hand edge of the disk, one polygon is chosen to be
the center of the projectile. Then  neighboring polygons are detected
(shown in black in color in the figure). These polygons are all given a
large initial velocity, directed towards the center of the disk, 
whose magnitude is obtained from the specified projectile energy. 

\begin{figure}
\begin{tabular}{cc}
\includegraphics[width=0.25\textwidth]{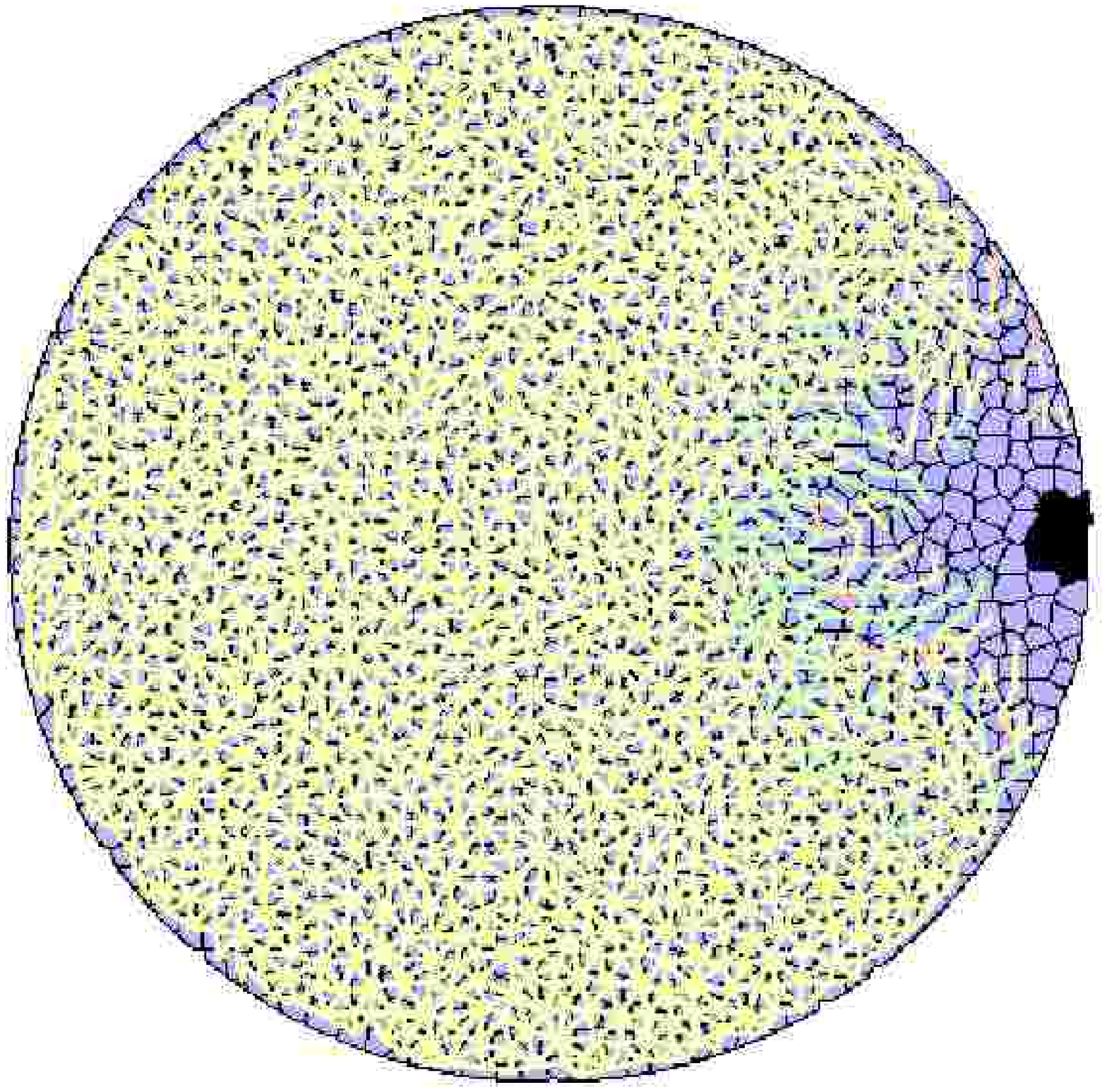}&
\includegraphics[width=0.25\textwidth]{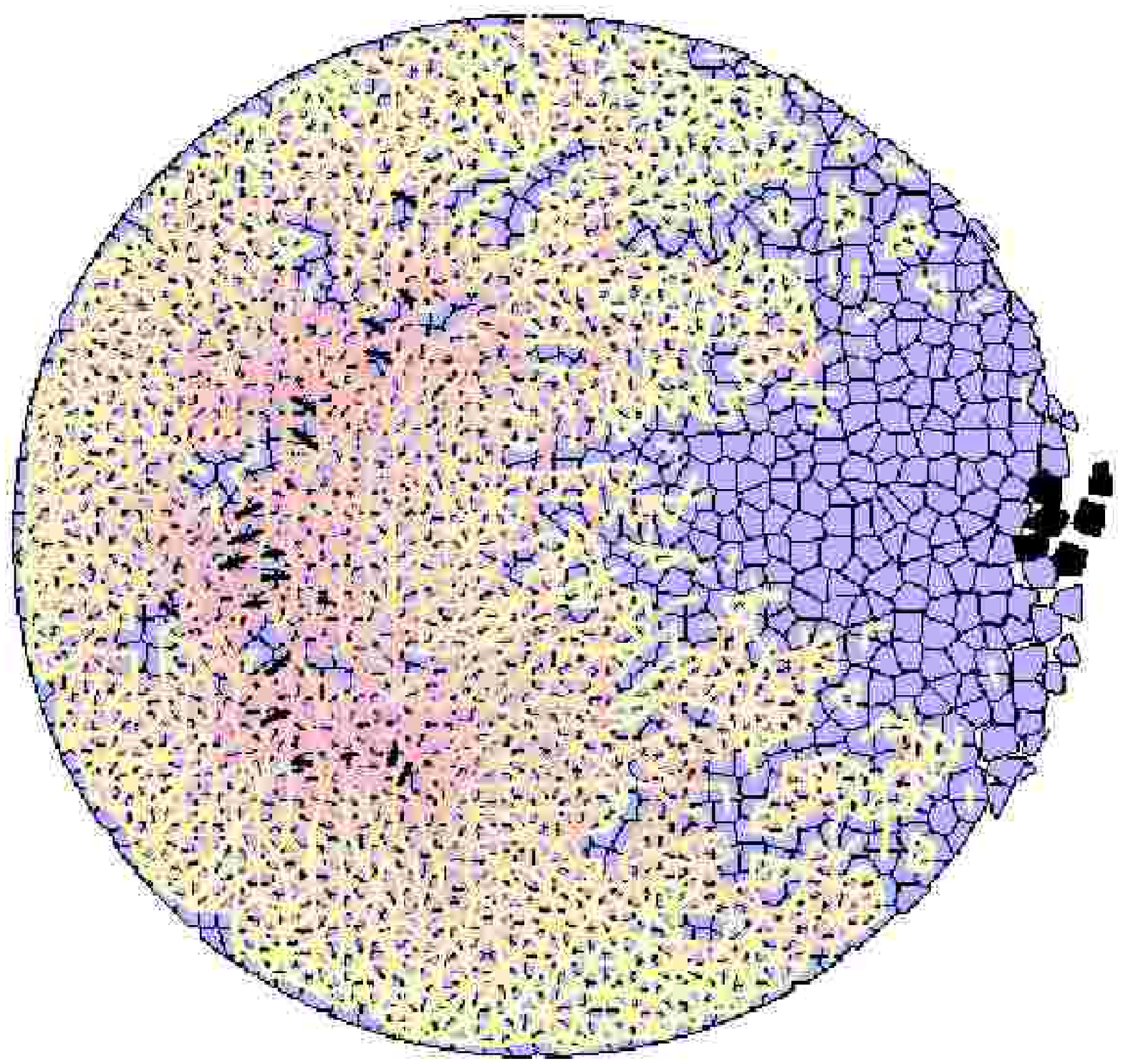}\\
(a)&(b)\\
\end{tabular}
\begin{tabular}{cc}
\includegraphics[width=0.26\textwidth]{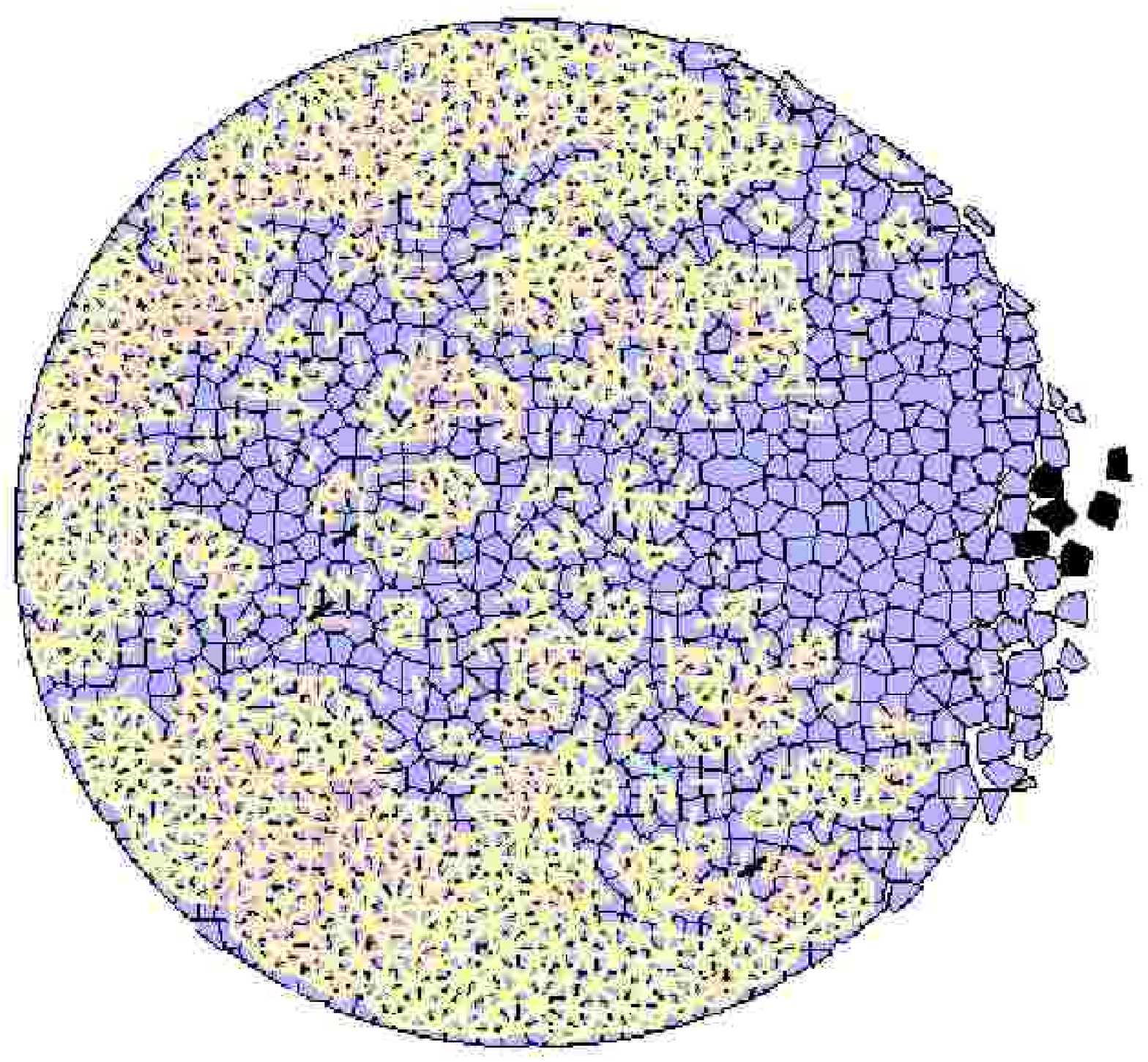}&
\includegraphics[width=0.27\textwidth]{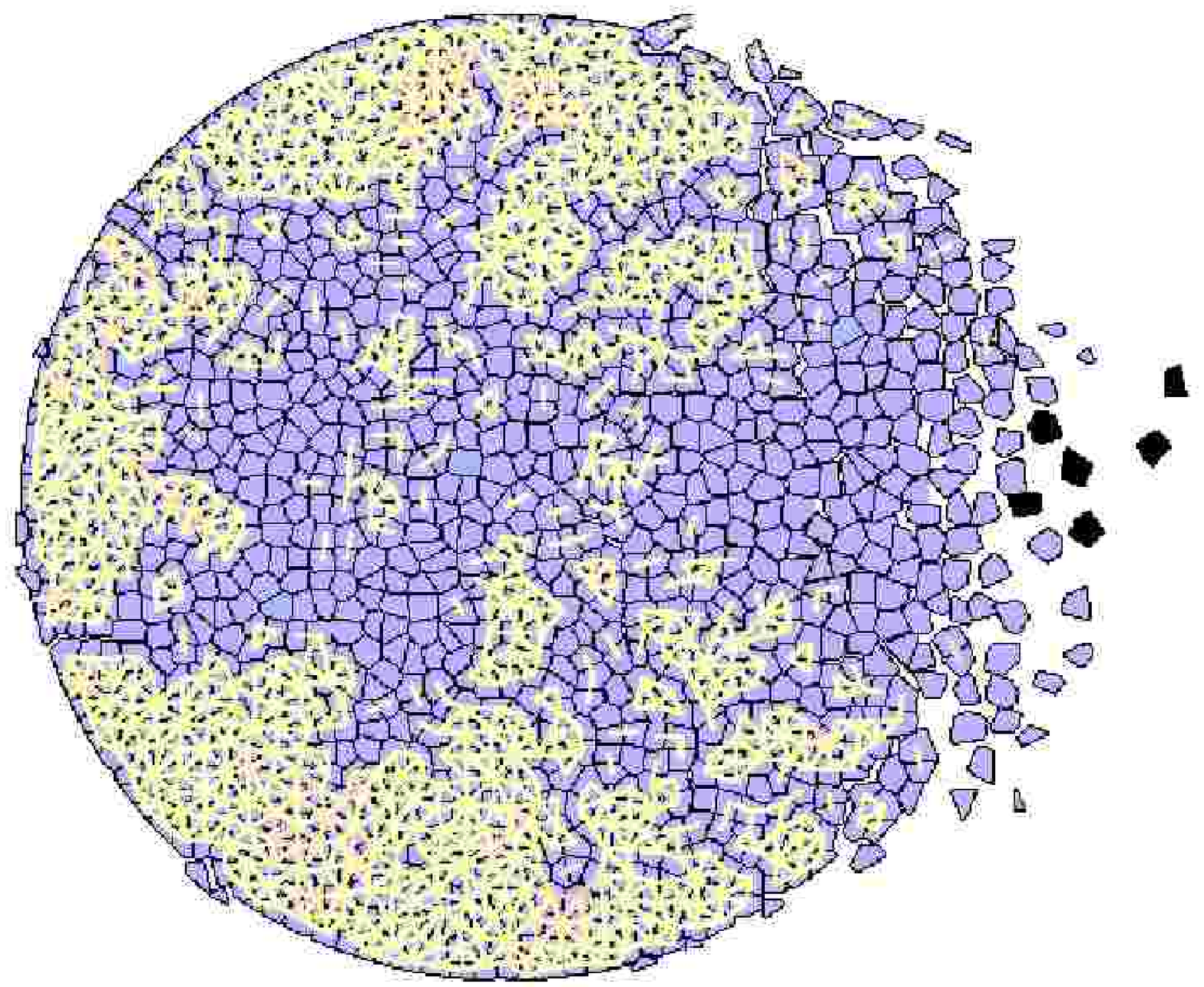}\\
(c)&(d)\\
\end{tabular}
\begin{tabular}{cc}
\includegraphics[width=0.25\textwidth]{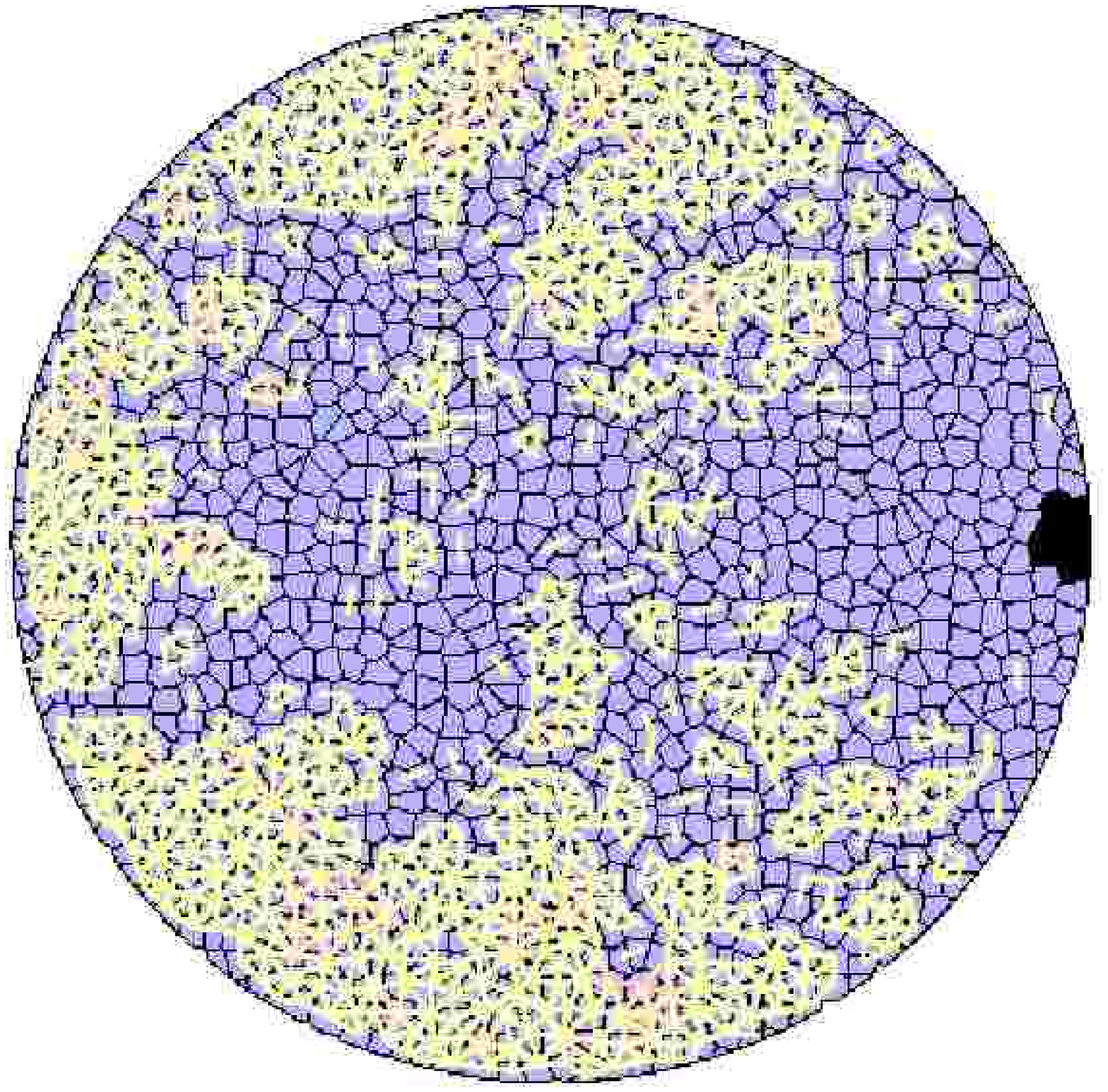}&
\includegraphics[width=0.25\textwidth, angle=90]{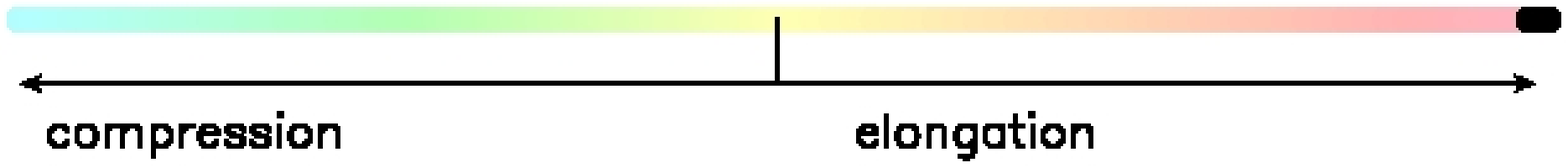}\\
(e)& \\
\end{tabular}
\caption{\label{snapshots}
Snapshots of the circular solid disc of radius 20 cm during
the fragmentation process at the critical energy $E_c=
1.6\times10^8$ erg. (a) After time 
0.00025 sec, the compression wave moving inwards creating the radial
cracks initiating from the impact point. (b) After time 0.00150 sec, the
compression wave reflected from free boundary surface and both compression
wave and elongation wave meet each other close to the boundary (c)
After time 0.00225 sec, a large number of beams break, forming cracks
perpendicular to the radial ones. (d) After time 0.00540, the shock wave
returns to the impact zone. (e) Fragments collected at the end of the
simulation and put back into their initial positions. The intact beams
are colored according to their longitudinal strain $\epsilon$.}
\end{figure}

\begin{figure}
\begin{tabular}{cc}
\includegraphics[width=0.23\textwidth]{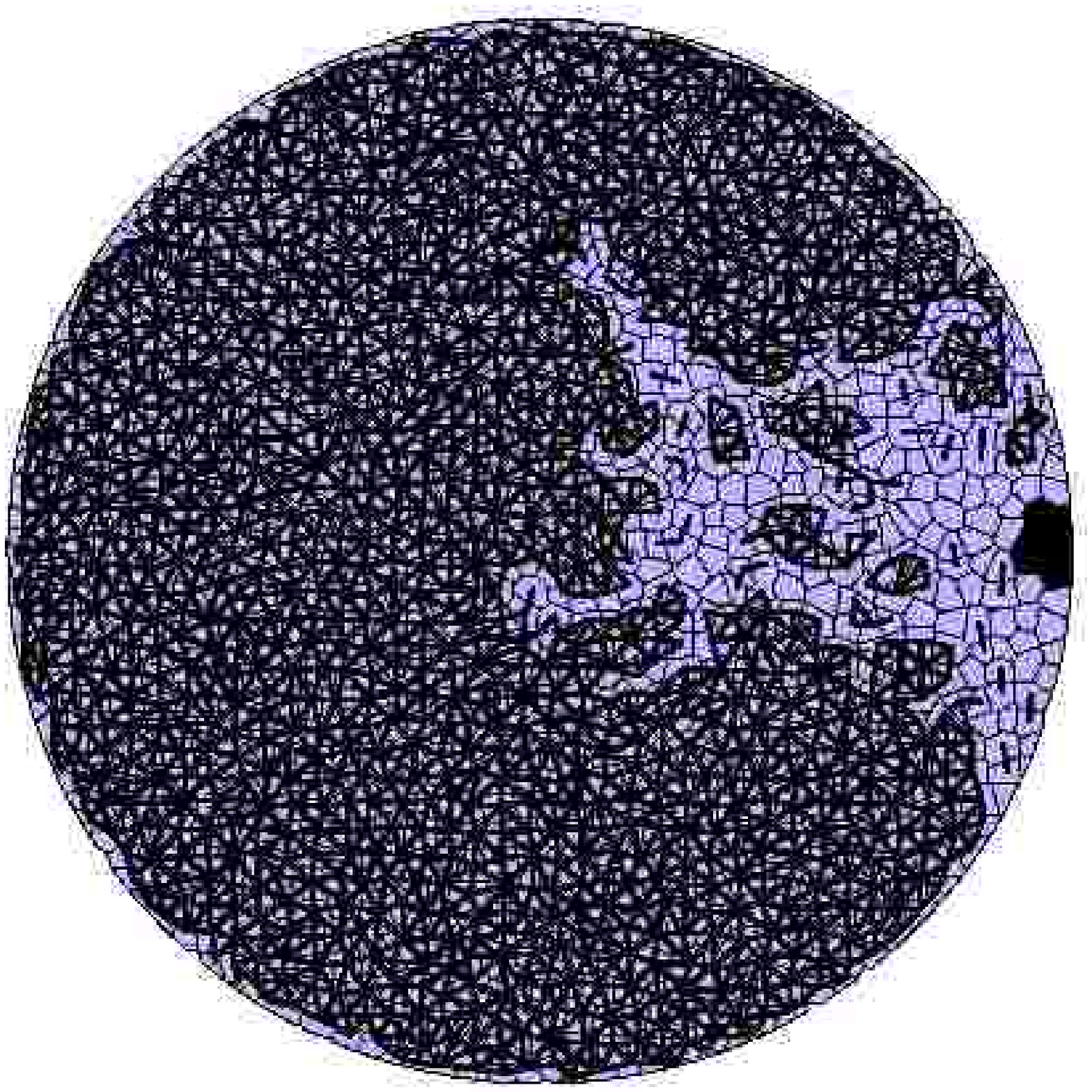}&
\includegraphics[width=0.23\textwidth]{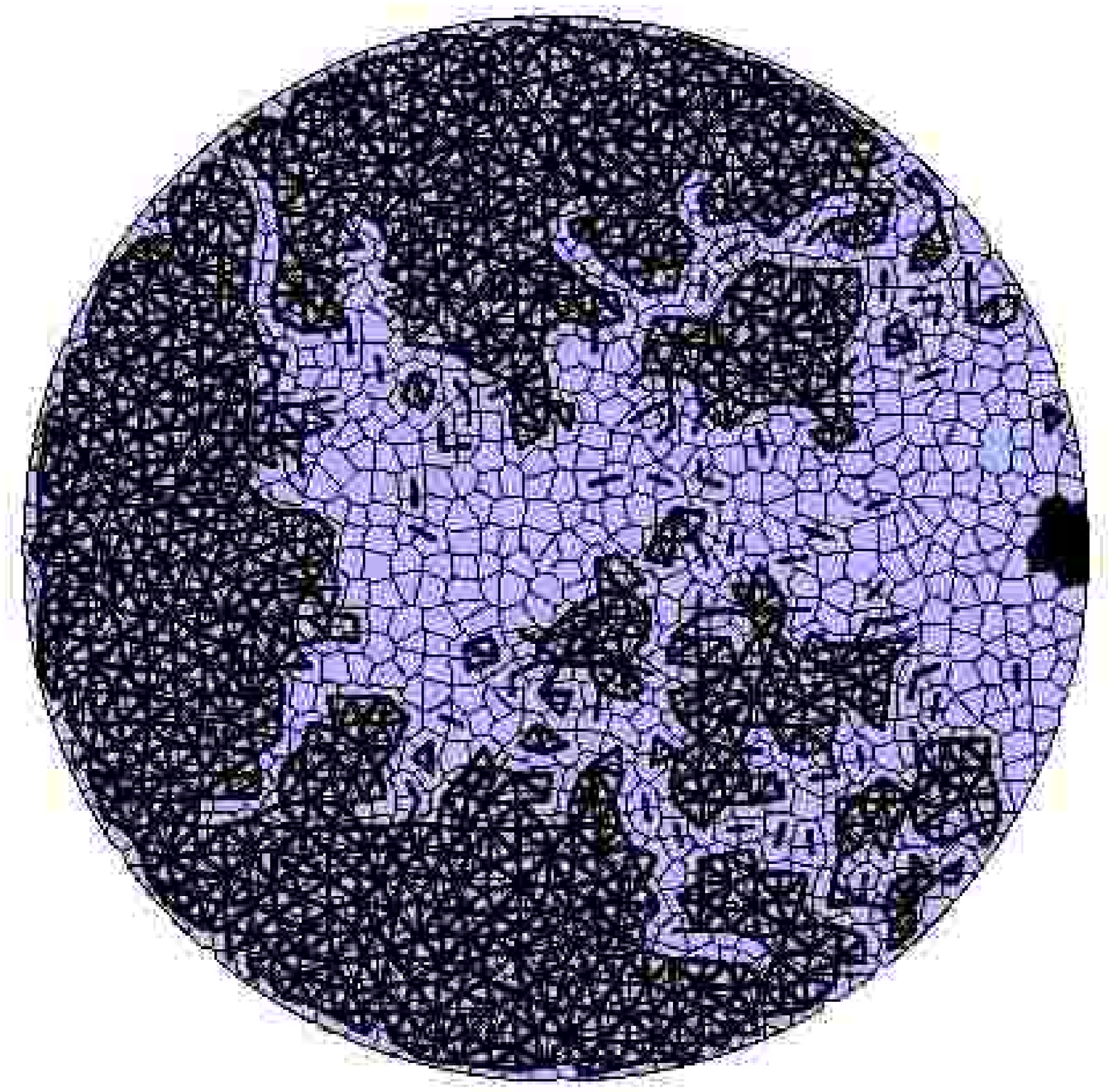}\\
(a)&(b)\\
\end{tabular}
\begin{center}
\begin{tabular}{c}
\includegraphics[width=0.23\textwidth]{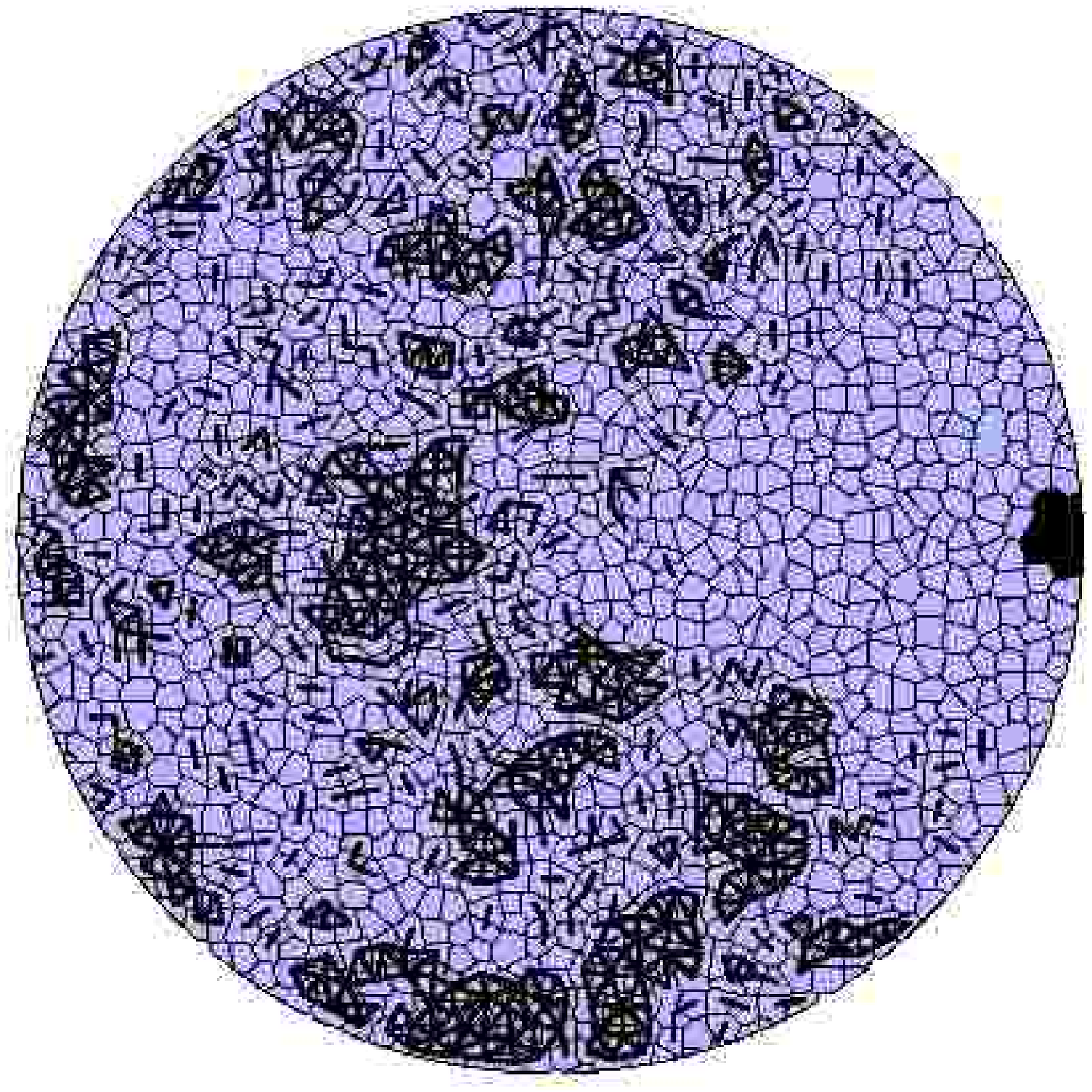}\\
(c)
\end{tabular}
\end{center}
\caption{\label{pattern}
The final reconstructed state of the crack pattern at the end of the 
fragmentation process obtained at different projectile energies
$E_o$. 
(a) $E_o = 0.5\times10^8$ erg,
(b) $E_o = 1.3\times10^8$ erg, 
(c) $E_o = 4.0\times10^8$ erg.} 
\end{figure}

In Fig.~\ref{snapshots} we present the fragmentation 
process obtained by simulation of a
disk of radius $20$ cm and a projectile energy of $1.6\times10^8$
erg. When the disk is struck by the projectile, a high
compression wave is formed at the impact site, where practically all the
beams are broken and all the fragments are single polygons. As the
compression wave moves towards the center of mass, radial cracks form
initiating from the impact point (see Fig.~\ref{snapshots}a).
The amplitude of the
initial compressive elastic pulse remaining after primary damage
strongly depends on the initial imparted energy and the amount of primary
damage that occurred in the contact zone. The pulse reflects at the
free boundary with opposite phase generating an incoming elongation
wave (Fig.\ \ref{snapshots}b).
A high stretched zone appears near the opposite boundary, which gives
rise to high breakage of beams and the propagation of cracks oriented
perpendicularly to the radial ones, Fig.~\ref{snapshots}c \cite{discrete}.
When almost all beams are
broken in this zone, the incoming elongation wave moves further towards
the impact point while simultaneously extending the cracks perpendicular
to the radial ones. When the wave reaches the impact zone,
the remaining energy is converted to kinetic energy of
the fragments. So, at the end of the simulation many single polygons are
flying away, back towards the origin of the projectile
(Fig.~\ref{snapshots}d). At the
end of the simulation we collected all the debris and reconstructed the
disc to investigate the final crack pattern, see Fig.~\ref{snapshots}e.

The final reconstructed state of crack pattern at the end of the 
fragmentation process obtained at different projectile energies is shown in 
Fig.~\ref{pattern}. It is observed that at very low energy (Fig.\
~\ref{pattern}a)  
mostly radial cracks are present which are the consequence of the
expansion of the sample perpendicular to the direction of the
projectile. In the energy range where solely radial cracks occur the
disc practically keeps its integrity and suffers only damage in the
form of cracks. Transverse cracks appear when the imparted energy
approaches a certain critical value $E_c$ (Fig.\ \ref{pattern}b).
At very high projectile energy (Fig.\ ~\ref{pattern}c), a large number
of radial  
as well as transverse cracks form and the disc breaks into 
numerous smaller pieces. 

Finally, two kinds of crack patterns are recognized: radial cracks
initiating from the impact point and cracks perpendicular to the radial
ones. The radial cracks developed prior to the perpendicular ones. The
crack propagation and the final crack pattern obtained by simulations
are consistent with the experimental
results \cite{platelike,glassplate}.

\section{Results}
\label{results}
\subsection{Transition from damage to fragmentation}

\begin{figure}
\begin{center}
\includegraphics[width=0.45\textwidth]{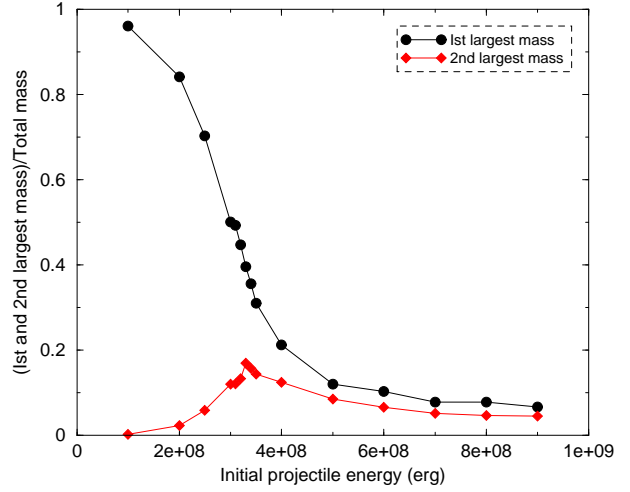}
\end{center}
\caption{\label{1st2nd} The first and second largest mass
as a function of the initial
projectile energy for a system size $R=30$ cm. The curve of the second
largest mass shows a peak corresponding to
 a sharp transition at a critical energy.}
\end{figure}

Previous studies on fragmentation identified a transition from damage,
where the target 
maintains its identity, to fragmentation, where it is broken into many 
small pieces.  The existence of these two states in the impact
fragmentation of discs is confirmed by
Fig.~\ref{1st2nd} where the mass of the largest and second largest
fragment normalized by the total mass are plotted as a function of the
projectile energy $E_o$.  At low energies the largest fragment contains
nearly all the original mass, and all other fragments are much smaller.
Thus, there are two classes of fragments: the largest mass is the
almost intact target, and the other fragments are small pieces that
were chipped off from the target by the projectile in the vicinity of
the impact point. The target has only been damaged, not destroyed.
At high energy however, the two largest fragments have comparable
masses and they are much smaller than the total mass indicating a
complete break up of the disc. In this energy range the target has
been entirely destroyed by the projectile.  

The transition between the two states occurs at a critical energy
$E_c$, which can be precisely defined as the energy that maximizes the
second largest fragment. Its value was determined $E_c\approx1.6\times
10^8$ erg for the system size $R=20$cm,
and $E_c\approx 3.3\times10^8$ erg for $R=30$cm. It is
important to note that the curve of the
mass of the largest fragment has a curvature change from convex to
concave, the position of which coincides with the maximum of the
second largest mass, {\it i.e.} with the critical energy $E_c$ .

\begin{figure}
\begin{tabular}{c}
\includegraphics[width=0.45\textwidth]{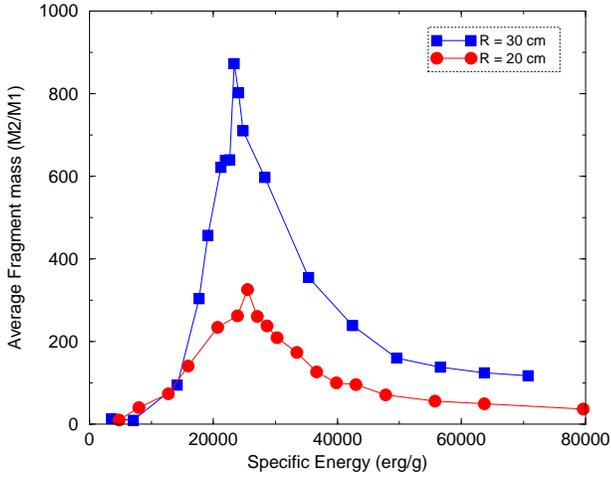}\\
\end{tabular}
\caption{\label{Mbar}
$M_2/M_1$ defined by Eq.\ (\ref{eq:m_k}) as a function of the specific
 projectile energy (projectile energy divided by the total target mass). 
Two distinct phases can be observed, the damaged and the fragmented one,
separated by a sharp transition. The critical energy was obtained as
$E_c = 1.6\times 10^8$ erg for the system size $R= 20$cm and $E_c=
3.3\times10^8$erg for $R=30$cm. The transition 
point is practically independent of system size when  $M_2/M_1$ is
plotted as a function of the specific energy.}
\end{figure}

To analyze in more detail the final state of the fragmentation
process, we evaluated the so called {\it single event moments} $M_k$
of fragment masses defined as
\begin{equation}
\label{eq:m_k}
M_k=\sum_i^N m_i^k-m_\mathrm{max}^k,
\end{equation}
where $N$ denotes the total number of fragments and $m_\mathrm{max}$ is
the largest fragment mass. We evaluated the average fragment
mass $\overline M$ defined as the ratio of the second and first moments
$\overline M \equiv M_2/M_1$. In Fig.~\ref{Mbar} $\overline M$ is
plotted as a function of the initial projectile energy $E_o$ for
system sizes $R=20$cm and $R=30$cm, where each point represents an average
over 20 simulations. 
The system again shows two distinct regimes, {\it i.e.}, damaged and
fragmented states, separated by a sharp maximum of $\overline M$
at the critical energy $E_c$. It is important to note that the
transition point is 
nearly independent of the system size $R$ when plotted as a function of the
specific energy, {\it i.e.}, the ratio of initial projectile energy to
initial total mass (see Fig.~\ref{Mbar} a). It can also be observed
that  increasing 
the system size makes the peak of $\overline M$ sharper, {\it i.e.}
the width of the peak decreases while the height increases, which is
typical for continuous phase transitions.

\begin{figure}
\begin{center}
\includegraphics[width=0.45\textwidth]{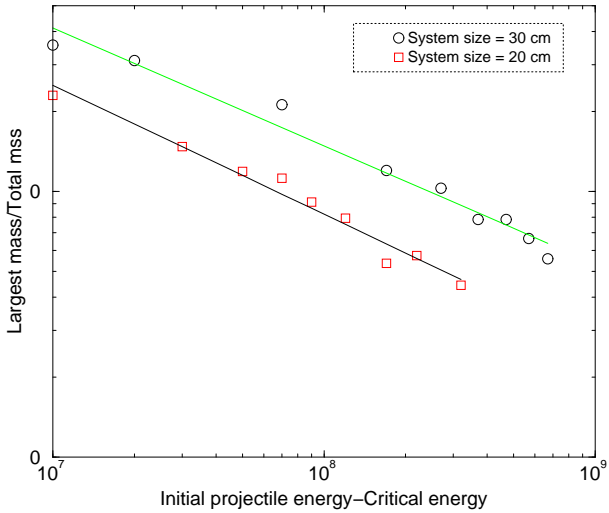}
\end{center}
\caption{\label{orderp}
The largest fragment mass divided by the total mass as a function of
the distance from the critical point $|E_o-E_c|$ for $ E_{o} >
E_c$. The exponent $\beta$ can be obtained as 
the slope of the least square fitted straight line.}
\end{figure} 

In order to characterize the behavior of the fragmenting system in
the vicinity of the critical point we plotted the normalized mass of
the largest fragment $m_\mathrm{max}/m_{tot}$
(where $m_{tot}$ is the mass of the target) obtained at different imparted
energies $E_o$ as a function of the distance
from the critical point $|E_o-E_c|$ for $E_o > E_c$. The value of
$E_c$ was varied until 
a straight line was obtained on a double logarithmic plot which also
provided an independent way of determination of $E_c$ from the previous
ones. Fig.\ \ref{orderp} shows that the functional form
\begin{equation}
\frac{m_\mathrm{max}}{m_\mathrm{tot}} \sim  |{E_{o} - E_c}|^{-\beta}
\quad E_{o} >  E_c ,
\end{equation}
is obtained above the critical point with a value of $E_c$ practically
coinciding with the value determined previously.
The value of the exponent $\beta$ was obtained as $\beta=0.49$ for
$R=20$cm and $\beta= 0.45$ for $R=30$cm. The two values are almost 
 the same with a precision of $\pm0.05$ implying that $\beta$ is
 independent of the system size.
\subsection{The Fragment Mass Distribution}

\begin{figure}
\begin{tabular}{cc}
\includegraphics[width=0.47\textwidth]{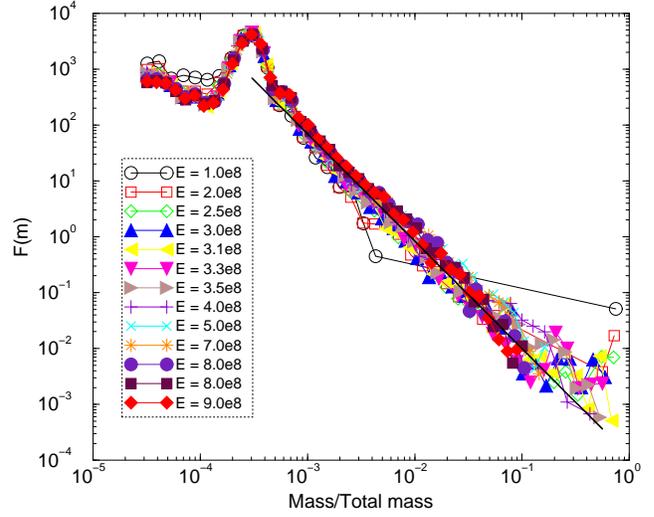}
\end{tabular}
\caption{\label{massdist}
The fragment mass histograms for the system size $R = 30$cm
with varying initial projectile energy. The straight line shows the power
law fitted to the curve at the critical energy $E_c = 1.6\times 10^8$ erg
with an exponent $\tau = 1.93$.}
\end{figure}

The fragment mass histograms $F(m)$ are presented in Fig.~\ref{massdist}
for the system size $R=30$ cm at varying initial
projectile energy $E_o$. $F$ is a probability density function so that
$\int F(m) dm=1$ its integral is unity.
In order to resolve the shape of the distributions at a wide range of
values, logarithmic binning was used, {\it i.e.}, the binning is
equidistant on logarithmic scale. The histograms have a maximum at
small fragment sizes due to the existence of single unbreakable polygons.  
Below the critical energy $E_{o} < E_c$ the distributions have a peak
at large fragments which gradually disappears when approaching the
critical point due to the break up of large damaged pieces into
smaller fragments. 
At the critical point $F(m)$ becomes asymptotically a power law over
almost three orders of magnitude of $m$. For the two system sizes studied 
the exponents are found to be practically equal and close to $2.0$,
consistent with previous numerical works
\cite{discrete,transition,twodisc,granulate}. 
More precisely, one obtains $\tau=1.93$ and $\tau=1.92$ for $R=20$ cm
and $R=30$cm, respectively.
Above the critical point  slight
deviations from the power law can be observed in the regime of large
fragments due to the finite size of the system. Since the 
effect is rather weak no definite form of this cut-off function could
be deduced, while in other types of fragmenting systems an
exponential cut-off function was obtained \cite{branching,univer,eggs}.

\subsection{Velocity distribution}

\begin{figure}
\includegraphics[width=0.45\textwidth]{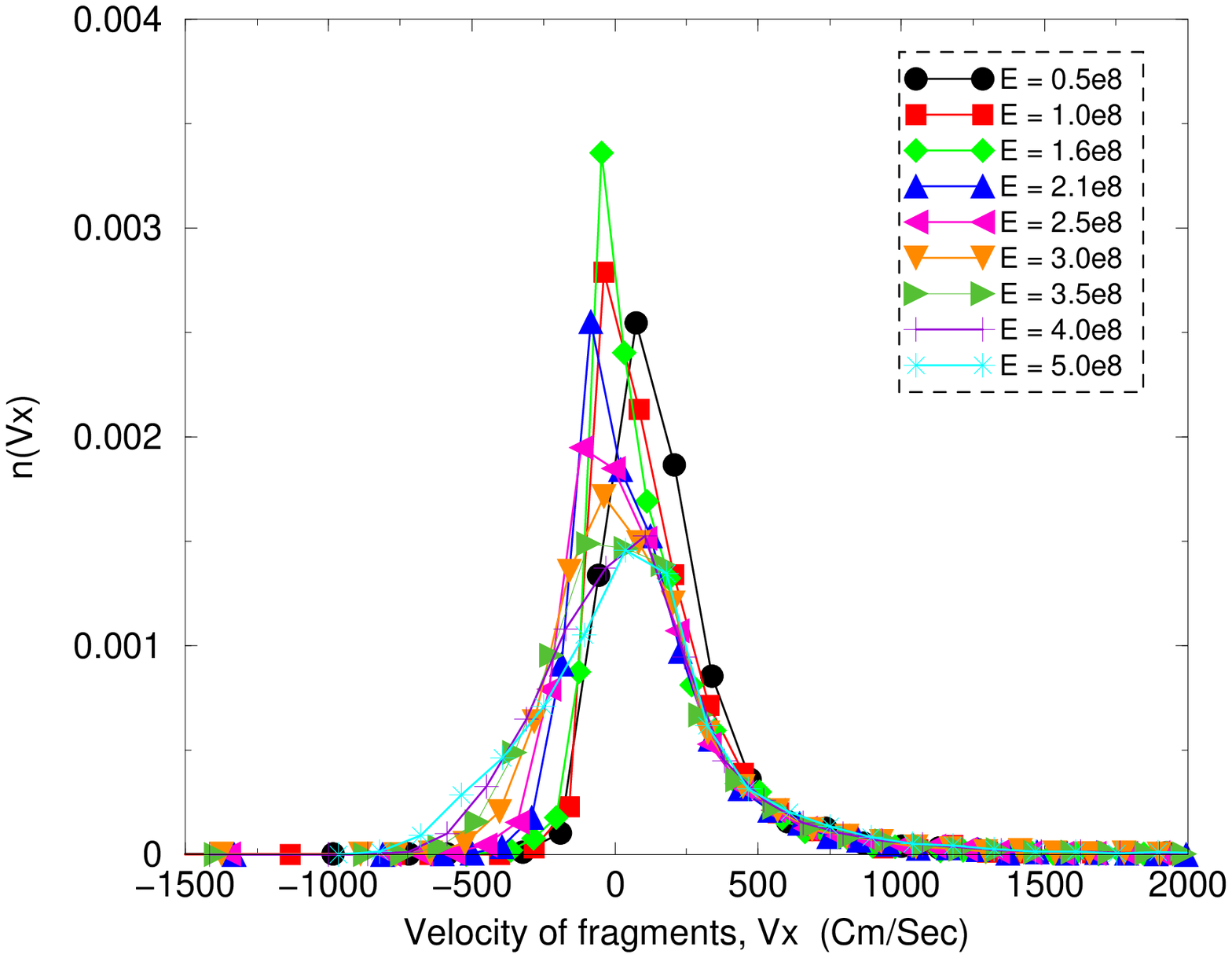}
\includegraphics[width=0.45\textwidth]{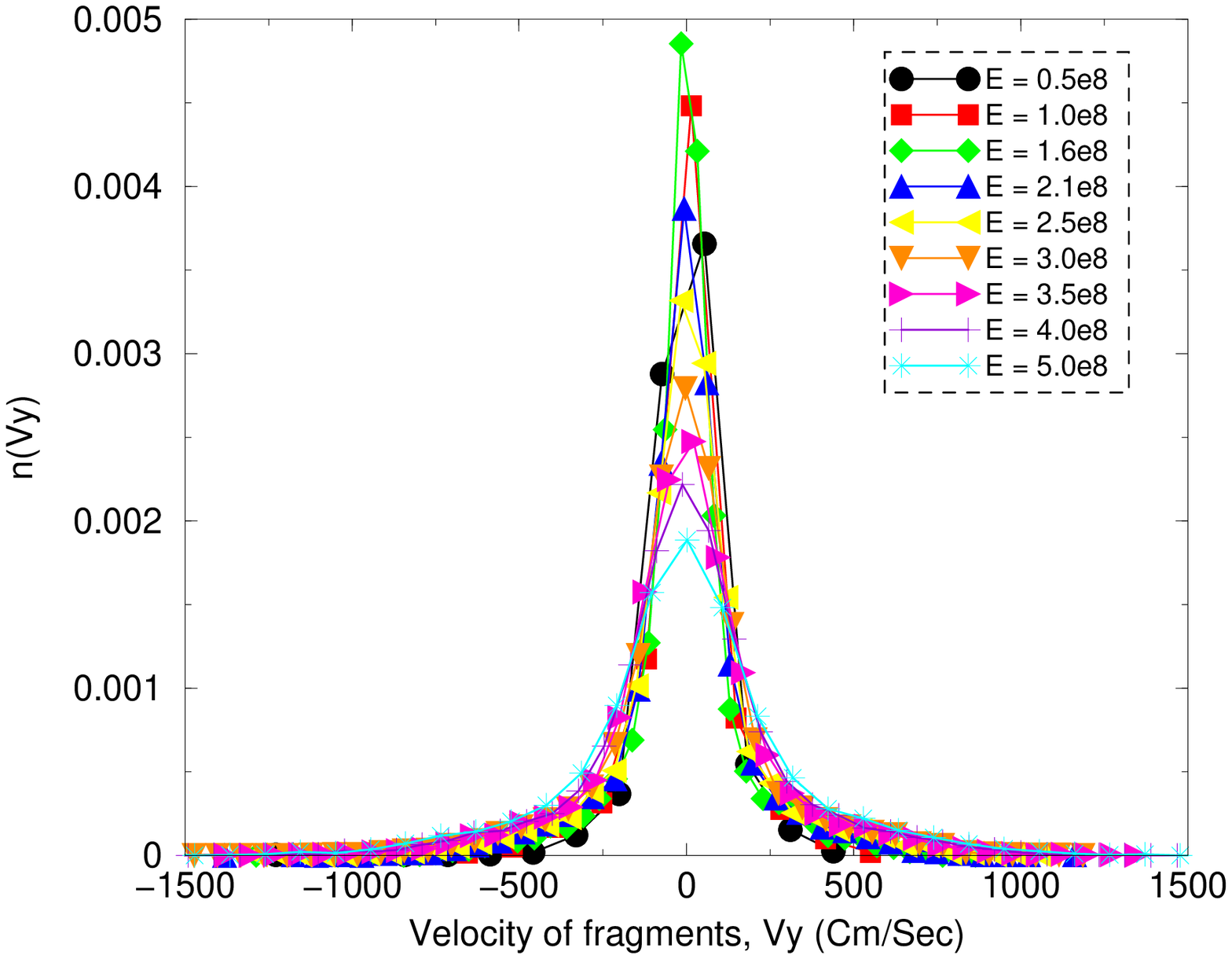}
\caption{\label{veldist}
The distribution of $x$ and $y$ components of the velocity of the
fragments for system size $R=20$ cm, for various
projectile energies.  Each curve represents an average over $20$
simulations.}
\end{figure}

In Fig.~\ref{veldist} the probability distribution functions
$n(v_x)$ and $n(v_y)$ of the
$x$- and $y$-components of the final fragment velocities are
plotted.  The initial projectile velocity is in the negative
$x$ direction.  One surprising feature of the $x$-component of the
velocities is that the majority of fragments are moving in 
the positive $x$ direction, {\it i.e.}, opposite to the initial direction
of the projectile, and therefore opposite to the total momentum
of the system.  This probably occurs because numerous high speed fragments
are ejected from the impact site, as shown in Fig.~\ref{snapshots}d. 
To satisfy momentum conservation,
the fewer, more massive fragments must be moving in the opposite
direction.  Thus, mass and velocity are correlated, which is
important to remember when we consider the energy distributions.

The distribution of the $y$-component of the velocity is symmetric
about $v_y = 0$ as expected from the symmetry of the initial conditions.
As the energy increases, the distribution broadens. 
\begin{figure}
\includegraphics[width=0.46\textwidth]{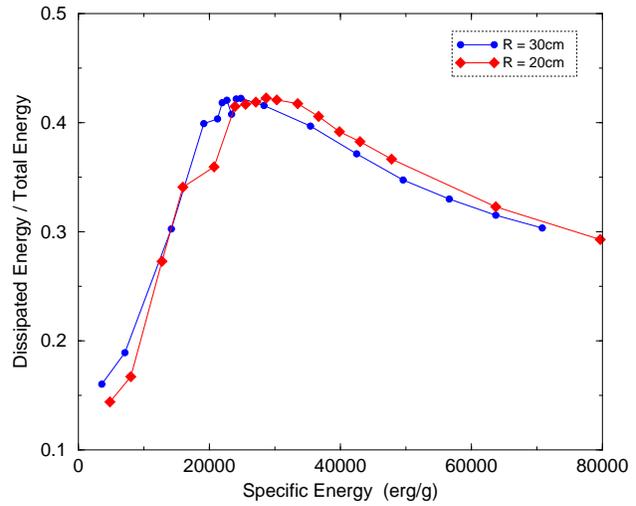}
\caption{\label{dissip}The ratio of the energy dissipated by cracking $E_d$
  and of the imparted energy $E_o$ as a function of the specific energy
  $E_o/m_{tot}$.} 
\end{figure}
It is worth noting that the maximum of $n(v_y)$ is largest when
the projectile energy is equal to $E_c$,
which implies that many fragments are formed with low 
kinetic energy at the critical point.
Hence, at the critical energy the efficiency of fragmentation is highest
because the largest fraction of the imparted energy is used for cracking.
Similar behavior is also observed in case of $n(v_x)$ except that
the maximum moves away from $v_x=0$  when the input energy 
 approaches the critical energy from below, and then 
moves back towards $v_x=0$
 as the projectile energy increases because the intensity of the 
elongation wave increases gradually beyond the critical point. 

\subsection{Fragment energies}
The efficiency of fragmentation can be quantified by the ratio of the
energy dissipated by cracking $E_d$ to the imparted energy $E_o$.
In Fig.\ \ref{dissip} the ratio $E_d/E_o$ is plotted as a
function of the specific energy $E_o/m_{tot}$. Obviously, the number of
broken bonds, and hence, 
$E_d$ is a monotonically increasing function of
the imparted energy $E_o$, however, the ratio $E_d/E_d$ shows a
maximum at the critical energy $E_c$ independent of the
system size, see Fig.\ \ref{dissip}. The highest fraction of $E_o$
used for cracking at the critical point can be estimated in the figure
to be $\approx 0.42$. 

\begin{figure}
\begin{center}
\includegraphics[width=0.46\textwidth]{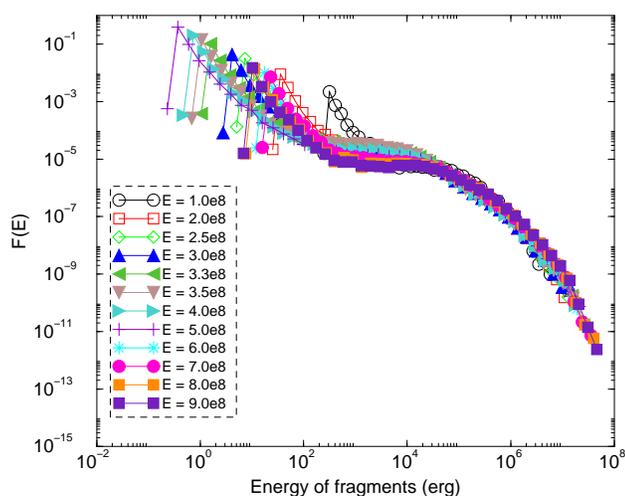}
\end{center}
\caption{\label{Edist}
Energy distribution of fragments at different projectile energies
for a disc of radius $R=30$cm.}
\end{figure}

In Fig.~\ref{Edist}, we show the energy distribution of the fragments.
Unlike the mass distributions, there is no clear power law form.  At large
energies, it seems that $F(E)\sim E^{-2}$ holds, but a closer examination
shows that the graphs are curved, not straight, in this region.

\section{Conclusions}

Motivated by recent experimental findings \cite{platelike,glassplate},
we have simulated the fragmentation of a disk by lateral impact of 
a projectile. The target was modeled as an assembly of unbreakable,
undeformable polygons bound together by elastic beams.
 The simulation permits us to follow the fragmentation
process in detail, allowing us to explain the crack patterns observed
experimentally \cite{platelike,glassplate}.
As the projectile energy increases a transition  \cite{transition}  
occurs from {\sl damage}, where the target is not destroyed, to {\sl
  fragmentation}, 
where the target is broken into many pieces.  The transition clearly
shows up in the behavior of the largest two fragments and the
average fragment mass as a function of the imparted energy.
We also found that the mass distribution $F(m)$ obeys a power law
over several orders of magnitude, with an exponent close to $2$,
in agreement with previous numerical results
\cite{discrete,transition,twodisc,granulate} and experimental findings
\cite{ice,composite,proton,platelike,glassplate,Au} on the
fragmentation of two-dimensional objects.  On the other 
hand, the energy distribution shows no such power law. 

As far as the design of comminution processes is concerned, this
work points out the importance of identifying the critical energy
$E_c$ that separates damage and fragmentation.  To maximize efficiency,
such processes should be designed to deliver energies just above $E_c$.
If the energy is too low, the target must be struck several times
in order to achieve a significant size reduction.  If the energy is too
high, much of it will be wasted, as size reduction does not increase
much above the critical energy.  Size reduction is usually not done
with high speed projectiles, but rather through collisions with hard
balls or with walls.  Therefore, consideration of these circumstances
is important.

\end{document}